\def\year{2022}\relax
\documentclass[letterpaper]{article} 
\usepackage{aaai22}  
\usepackage{times}  
\usepackage{helvet}  
\usepackage{courier}  
\usepackage[hyphens]{url}  
\usepackage{graphicx} 
\usepackage{natbib}  
\usepackage{caption} 
\DeclareCaptionStyle{ruled}{labelfont=normalfont,labelsep=colon,strut=off} 
\usepackage{algorithm}
\usepackage{algorithmic}

\usepackage{newfloat}
\usepackage{listings}
\usepackage{dirtytalk}
\usepackage{booktabs}
\usepackage{subcaption}
\usepackage{array}
\floatstyle{ruled}
\newfloat{listing}{tb}{lst}{}
\floatname{listing}{Listing}

\newcommand{\eg}{\emph{e.g.}}
\newcommand{\ie}{\emph{i.e.}}

\title{AI-Cybersecurity Education Through Designing \\AI-based Cyberharassment Detection Lab}
\author {
     Ebuka Okpala\textsuperscript{\rm 1}, Nishant Vishwamitra\textsuperscript{\rm 2}, Keyan Guo\textsuperscript{\rm 3}, Song Liao\textsuperscript{\rm 1}, Long Cheng\textsuperscript{\rm 1}, Hongxin Hu\textsuperscript{\rm 3}, Xiaohong Yuan\textsuperscript{\rm 4}, Jeannette Wade\textsuperscript{\rm 5} and Sajad Khorsandroo\textsuperscript{\rm 4}
}

\affiliations {
    \textsuperscript{\rm 1} School of Computing, Clemson University\\
    \textsuperscript{\rm 2} Department of Information Systems and Cyber Security, University of Texas at San Antonio\\
    \textsuperscript{\rm 3} Department of Computer Science and Engineering, University at Buffalo\\
    \textsuperscript{\rm 4} Department of Computer Science, North Carolina A\&T State University \\
    \textsuperscript{\rm 5} University of North Carolina at Greensboro
\\
}

\begin{document}

\maketitle

\begin{abstract}
Cyberharassment is a critical, socially relevant cybersecurity problem because of the adverse effects it can have on targeted groups or individuals. While progress has been made in understanding cyber-harassment, its detection, attacks on artificial intelligence (AI) based cyberharassment systems, and the social problems in cyberharassment detectors, little has been done in designing experiential learning educational materials that engage students in this emerging social cybersecurity in the era of AI. Experiential learning opportunities are usually provided through capstone projects and engineering design courses in STEM programs such as computer science. While capstone projects are an excellent example of experiential learning, given the interdisciplinary nature of this emerging social cybersecurity problem, it can be challenging to use them to engage non-computing students without prior knowledge of AI. Because of this, we were motivated to develop a hands-on lab platform that provided experiential learning experiences to non-computing students with little or no background knowledge in AI and discussed the lessons learned in developing this lab. In this lab used by social science students at North Carolina A\&T State University across two semesters (spring and fall) in 2022, students are given a detailed lab manual and are to complete a set of well-detailed tasks. Through this process, students learn AI concepts and the application of AI for cyberharassment detection. Using pre- and post-surveys, we asked students to rate their knowledge or skills in AI and their understanding of the concepts learned. The results revealed that the students moderately understood the concepts of AI and cyberharassment.
Comparing the learning experiences of students in the Spring and Fall semesters using the post-surveys, in the Spring semester, while students understood the purpose of detecting cyberharassment, their knowledge of how AI works, state-of-the-art (SOTA) cyber-harassment detectors and automated cyberharassment detection did not improve. After refining the lab using the student feedback from the Spring semester, students' knowledge of AI and automated cyberharassment detection significantly improved. We learn that supplementing the hands-on lab with a theoretical background lecture improves understanding of the lab contents, indicating that supplementing experiential learning with theoretical constructs improves the student learning experience. These findings confirm that the developed lab is viable for teaching AI-driven socially relevant cybersecurity to non-computing students and can be used by other institutions.
\end{abstract}
\section{Introduction}
With the rise of social media, cyberharassment (\eg, cyberbullying and cyberhate) has become more prevalent in daily interactions~\cite{SoK:Harassment:2021}.
It often involves {\em inappropriate online behavior} and {\em deliberate cyber threats} against individuals (such as teenagers~\cite{cyberbullying}), or specific social groups on the grounds of characteristics such as race, sexual orientation, gender, or religious affiliation~\cite{cyberhate}.
Cyberharassment is identified as a critical socially-relevant cybersecurity problem~\cite{socialcybersecurity, socialcybersecurity2}, since it can have significant negative impacts on the safety and emotional well-being of targeted groups, especially teens and minority communities. The Cyberbullying Research Center's research reported that 37\% of middle and high school students have been cyberbullied during their lifetime~\cite{CyberbullyingStatistics}, and this number is expected to further increase as teens continue to have an increased online presence. Cyberharassment can even result in catastrophic consequences of increased suicide among the affected teens who are unable to appropriately get away from the harassment~\cite{Daniel:phdthesis:2017}. The shift from traditional text-based cyberharassment to {\em multimodal} (\ie, both texts and images)~\cite{Memes} cyberharassment poses a challenge to effective cyberharassment detection.

Artificial Intelligence~(AI)/Machine
Learning (ML) has immense potential to solve this critical problem.
Automatic detection methods of both text-based and image-based cyberbullying using AI techniques have emerged~\cite{Cyberbullying:IJCAI:2016}.
Internet companies such as Facebook and Google have also deployed AI algorithms to detect toxic content on social media~\cite{facebook, google}. 
Meanwhile, adversaries may exploit vulnerabilities of AI-based classifiers to evade existing cyberharassment detectors~\cite{DBLP:journals/corr/GoodfellowSS14, TextBugger:2019:NDSS, DBLP:journals/tist/ZhangSAL20}. There exist {\em social problems}, such as fairness and ethics, in AI models for cyberharassment detection. For example, some particular demographic groups are unfairly treated by AI-based detectors~\cite{sap-etal-2019-risk, okpala2022aaebert}. 
Concerns have been raised that the vulnerabilities of AI models as well as the robustness against attacks are biased towards underrepresented groups~\cite{DBLP:journals/corr/abs-2006-12621}. 
As such, an unfair AI-based cyberharassment detection system may perpetuate and aggravate existing prejudices and inequalities in society.

As cyberharassment grows online, particularly on social media, there is a need to equip computing students with the AI skills and knowledge required to design and develop AI-based systems to detect and remove cyberharassment. As the field of cyberharassment is interdisciplinary, to develop better detection systems, non-computing students, especially social science students, need to have a general understanding of AI and how it is being used in detecting cyberharassment. A major concern with University training is how siloed it can be and how challenging it is for young adults to truly explore their options. Interventions such as the one outlined here would enable effective collaboration between Computer science and social science students and researchers to create better cyberharassment detection systems. 
It is, therefore, imperative to teach AI-based cyberharassment in universities to both computing and non-computing students. This intervention also benefits social science students by providing insights into how they can address social problems using computer science, a discipline they may have otherwise had no exposure. This practical application of their social scientific passions may influence the degree minors or graduate programs students pursue moving forward as more researchers are needed in this intersectional area since this is a growing area of concern. 

To teach and engage students in learning cybersecurity and AI-related topics such as data science, instructors have adopted a wide range of pedagogical methods such as flipped classroom \cite{nwokeji2019effect, eybers2016teaching}, project-based learning \cite{nwokeji2017cross, velaj2022designing}, gamification \cite{matovu2022teaching}, among others. Experiential learning has been regarded as one of the best ways to train future engineers by engineering educators \cite{samavedham2012facilitating}. Towards this end, experiential learning could be used in teaching AI socially relevant cybersecurity to non-computing students. 

Experiential learning, simply put, is learning from experience or learning by doing. More formally, experiential learning is a type of active learning where students learn through experience \cite{kolb2005learning, kolb2014experiential}. Experiential learning is active rather than passive. Instructors have recognized how instrumental experiential learning is in providing students with valuable hands-on experience in an AI/ML-related field such as data science \cite{barman2022experiential, allen2021experiential, rosenthal2020data, anderson2014undergraduate}. 
In our study, we teach AI-based socially relevant cybersecurity for cyberharassment detection for two semesters using a hands-on experiential lab. Before the introduction of the lab, a questionnaire was used to rate the AI skills and knowledge of the students. After the end of the lab, another questionnaire was used to ask the students to rate their skills and knowledge of AI and AI-based cyberharassment detection covered in the lab. 

Our analysis and statistical results (using sample t-test) showed that experiential learning engages students in learning AI-based cyberharassment, and it is viable for teaching AI skills to non-computing students. Also, our findings show that having a theoretical lecture before the experiential lab improves understanding of the lab contents. These findings confirm that the developed lab is viable for teaching AI-driven socially relevant cybersecurity to non-computing students and can be used in other institutions.

\section{Related Work}
Instructors have mainly employed active learning paradigms such as experiential learning to teach AI/ML-related courses, mostly in engineering and Computer Science. The shortcoming of standard pedagogical methods in data science in online courses and data science specializations are detailed in \cite{serrano2017experiential}. Noting that experiential learning mitigates the shortcomings as it focuses on problems to be solved instead of on specific methods being used. Using the experiential learning style theory introduced in \cite{kolb2014experiential}, they developed a framework to create experiences in a deep learning course. Finally, they review dataset repositories used in data science and propose requirements for an experiential learning platform to offer experiences.  

Understanding the importance of capstone projects in data science courses, \cite{allen2021experiential} observed that more attention needs to be paid to how these projects or curriculum are structured. In their work, they develop an interdisciplinary, client-sponsored capstone program in data science and machine learning. In the program, students from different undergraduate and graduate degree programs engage in experiential learning by completing a large-scale data science or machine learning capstone project toward the program's end— the projects were framed to be challenging and encompass all aspects of data science. The curriculum was split into modules focusing on the data science pipeline, ethics, and communication. It was developed using evidence-based approaches from capstone and design programs in engineering, practicums, and other project-based courses. 

In \cite{barman2022experiential}, the challenges in using capstone projects as experiential learning opportunities in data science courses due to resource constraints and data legalities involved in students working with clients on clients' real-world data sets are emphasized. A novel client-facing consulting data science course that provides experiential learning to undergraduate and graduate students is developed to tackle this issue.

In an ethics and data analytics course where engineering students developed solutions for smart grid, smart health, and smart mobility, \cite{martin2022enacting} explored how professional responsibility is understood by engineering students working on a solution to a real-world problem proposed by a client. The authors acknowledge that as technology such as AI/ML advances, it presents complex challenges that require an interdisciplinary approach. In line with this, they introduce students to real-life problems presented as socio-technical challenges embedded in a learning context using Challenge Based Learning (CBL), an experiential learning method. 

The most recent works more closely related to our work are the works of \cite{salazar2022designing} and \cite{hashim2022first}. AI education is a challenging task because it is not well studied, making it one of the challenges in engineering education \cite{salazar2022designing}. With this knowledge, through a series of directives, acts, and laws, the United States of America government highlighted the importance of the Department of Defence to embrace AI at speed and scale \cite{salazar2022designing}. In \cite{salazar2022designing}, the Department of Defence (DoD) and the United States Air Force (USAF) partnered with MIT to design and develop educational research activities that will provide AI training for DoD and USAF personnel with diverse professional and educational backgrounds from high school to graduate degrees, and to the general public. The educational research activities explored various ways to teach AI education, online and in-person, to deliver experiential learning experiences to learners of diverse backgrounds. In the program's first iteration, they developed and studied the learning journeys of three different types of USAF employees (leaders, developers, and users). The findings will be used in future iterations of the program. 

The authors in \cite{velaj2022designing} design a data science course for non-computer science students. The course goal was to develop a new method for designing a data science course suitable for teaching students with different backgrounds in data science. The goal of the new method was to ensure that students gained skills on how to set up, manage, and conduct data science projects. The data science course was taught in Business Analytics and Data Science and the Digital Humanities master's programs. Students were introduced to KNIME, an open-source analytics platform for creating data science workflows without coding. Students were also encouraged to use Python and R programming languages if they were already familiar with them. The results reveal that the course is well designed and structured for students with different backgrounds, that the students gained skills to carry out a data science project, and that students liked the analytics platform used in the project work. 

The popularity of AI/ML has led to the proliferation of research studies on designing better data science education materials, as shown in some of the works reviewed. These works mainly focused on designing data science courses for engineering and computing students, with a few focusing on non-computing students. Most importantly, these works do not focus on experiential, hands-on labs that will provide students with the experience to bridge the gap between theoretical knowledge and practice. We fill this gap by developing experiential, hands-on labs for non-computing students and discuss the lessons learned in developing these hands-on labs. 
\section{Design \& Development of AI Socially Relevant Cyberharassment Lab}
\subsection{Lab Structure}
The experiential learning laboratory consists of two labs where students become familiar with the basics of AI and the AI/ML pipeline for applying ML to a problem.  
\subsubsection{Objectives}
We designed the experiential learning laboratories with specific learning objectives. Essential for guiding student learning, the learning objectives guided and helped develop our laboratories. This ensured that our study covered the fundamental elements of AI and different dimensions of AI-driven social cybersecurity. It also demonstrated the interplay between AI and cybersecurity and how AI is used for cyberharassment detection. Specifically, our learning objective is to develop hands-on experiential labs that will increase general awareness of socially relevant cybersecurity and AI, which is suitable for teaching AI socially relevant cybersecurity to non-computing students. 

\subsubsection{Lab}


Our lab adopts a phased design approach. Initially, AI and cybersecurity experts in our team designed the preliminary lab and implemented and integrated the cyberharassment detection code in the lab. The lab was designed considering the student profiles, AI and cyberharassment learning objectives, and desired skills. Subsequently, social scientists in our team engaged in dialogues with the AI and cybersecurity scholars to enhance the lab's reach and inclusivity. After the concerns of the social scientists have been addressed in the next iteration of the lab, the researchers collaboratively designed the lecture sessions and lab assignments. The collaborative and interdisciplinary approach ensured that the lab was accessible to a broad range of learners.

\begin{figure}[h]
    \centering
    \includegraphics[width=0.47\textwidth]{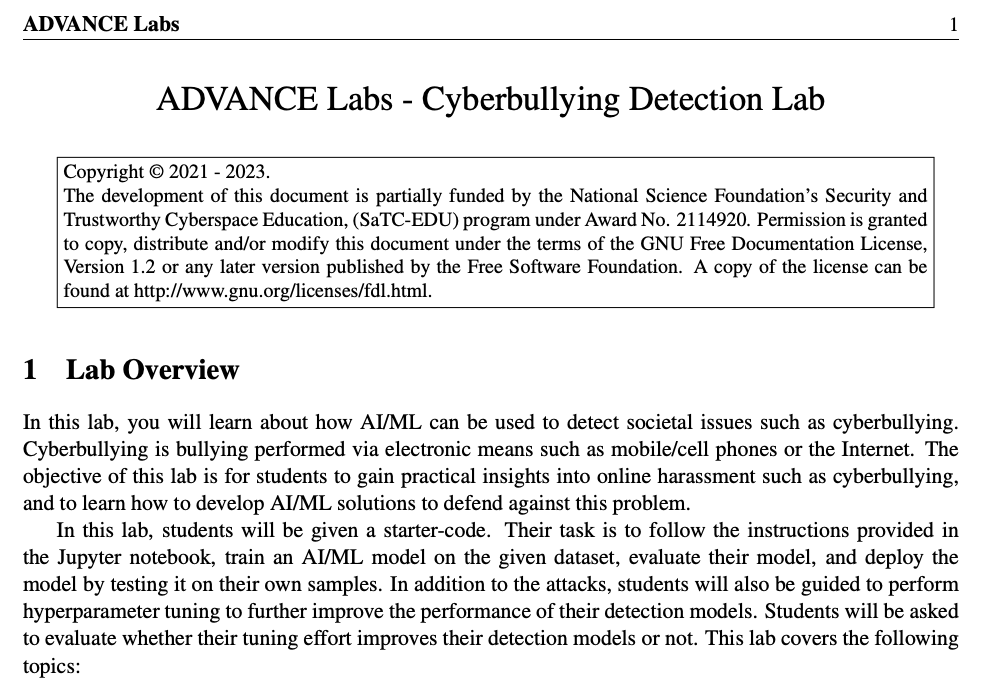}
    \caption{A screenshot illustrating Lab 1 instruction manual}
    \label{fig:lab1_manual}
\end{figure}

Our lab, \textit{Cyberbullying Detection Using AI}, is designed to guide students through a series of learning objectives and the AI development process. The learning objectives include understanding AI, understanding the concept and severity of cyberharassment, the importance of using AI in addressing this widespread online social issue, and introduction to the development and the use of AI systems for cyberharassment detection. For the AI development process, the AI experts ensured that the design followed the AI development pipeline (data collection, verification and preprocessing, feature extraction, AI system training, and testing, and use of trained AI systems on a task) to provide students with the fundamentals of the processes followed to develop an AI system. In parallel, the social scientists offered profound insights into the peculiarities of cyberharassment, highlighting its significance and the pressing need for AI intervention. All researchers from diverse academic backgrounds involved in this work cross-verified the lab content to ensure the lab is easily accessible to non-computing students. This process ensures that those new to AI and cyberharassment can easily grasp and engage with our lab materials. Figure~\ref{fig:lab1_manual} shows a comprehensive instruction manual prepared for the lab to facilitate independent learning and ensure students accomplish the learning objectives outside of the classroom.

\begin{figure}[t]
    \centering
    \includegraphics[width=0.5\textwidth]{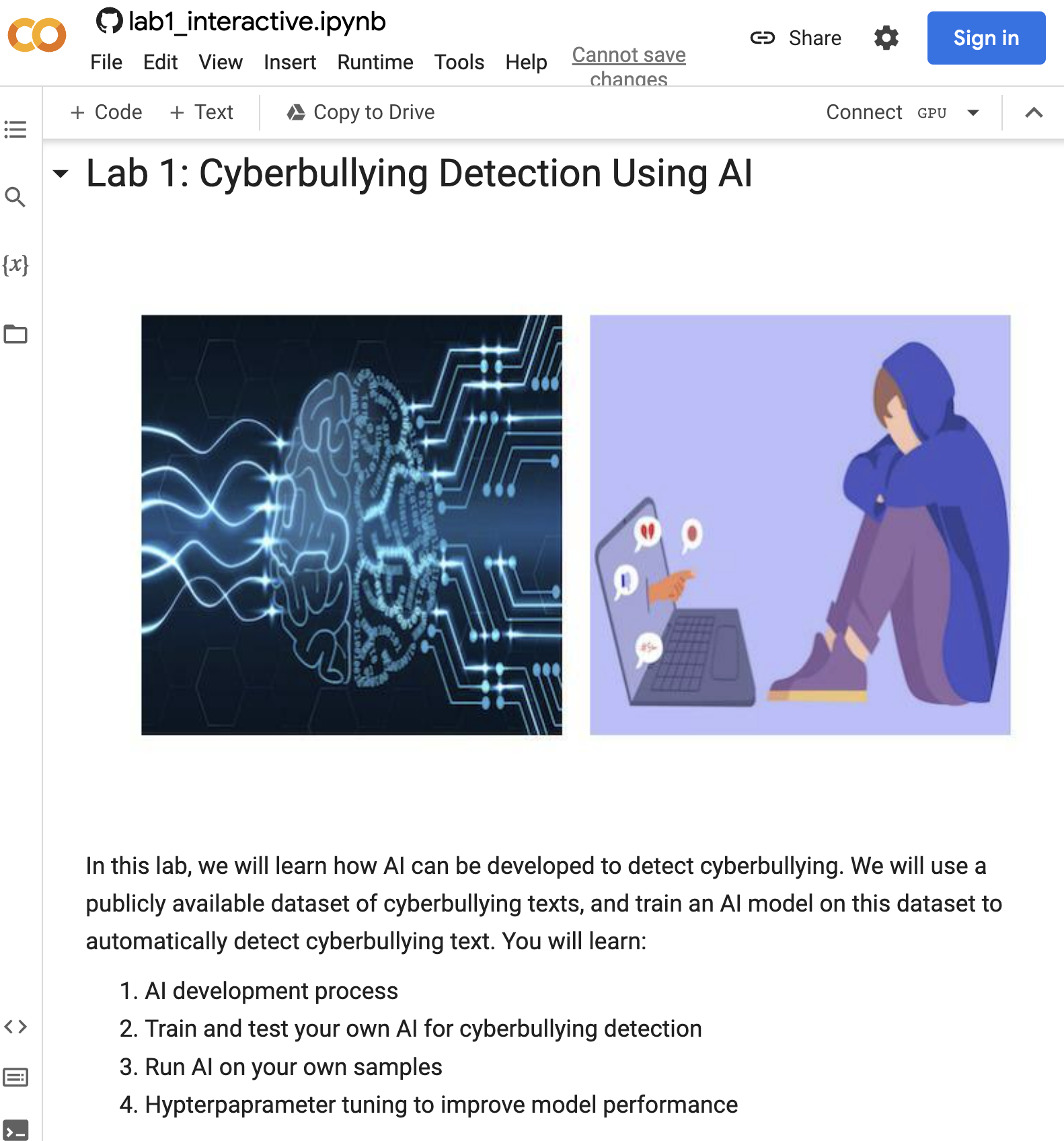}
    \caption{A screenshot illustrating the lab interface}
    \label{fig:lab1_interface}
\end{figure}

\subsubsection{Lab Delivery}
To facilitate a comprehensive understanding of our lab, we implement a three-fold approach that includes a lecture phase, an experiential hands-on experience phase, and a phase dedicated to independent work.


Before the students can work on the hands-on lab, a background lecture designed and developed by the team of researchers is given to the students. In the lecture, the students get acquainted with the nature and nuances of cyberharassment, AI, the AI development process, and the need for utilizing AI for cyberharassment detection.

\begin{figure}
    \centering
    \includegraphics[width=0.48\textwidth]{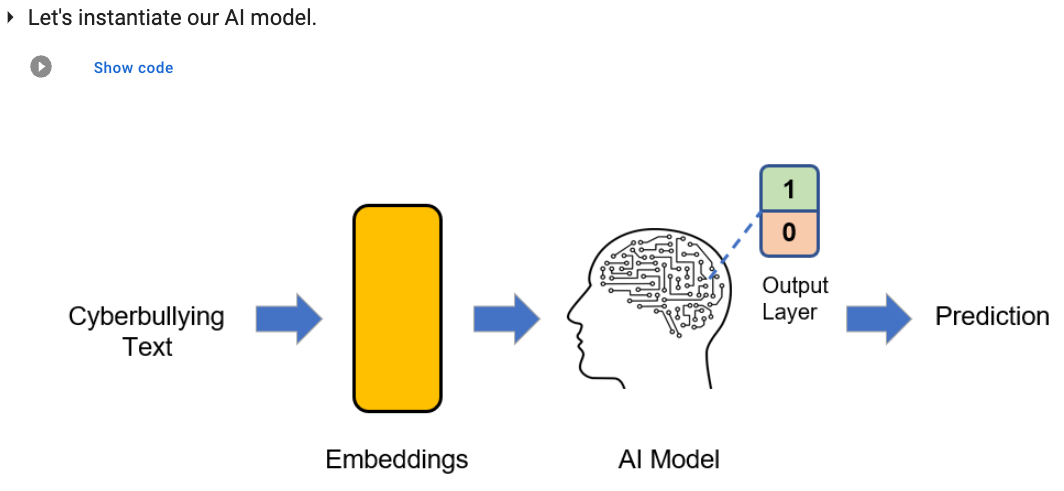}
    \caption{The visual aid for model explanation in Lab 1}
    \label{fig:model}
\end{figure}

After introducing the students to the fundamentals needed to complete the lab through the lecture, students are introduced to the hands-on lab and guided through the hands-on experience platform depicted in Figure~\ref{fig:lab1_interface}. The lab is developed on the Google Colab platform. The hands-on experience allows students to apply theory in practice. It facilitates a deeper understanding of how AI solutions are developed and, most importantly, how AI can be utilized to mitigate cyberharassment.
We have chosen to utilize the Google Colab platform for several compelling reasons. First and foremost, Google Colab provides an interactive environment that integrates text and code cells. This not only enables us to write and execute Python code for deploying AI models and detecting cyberharassment content, but it also allows us to provide clear, step-by-step explanations alongside the code. Moreover, these text cells can be utilized to embed visual aids and explanations such as Figure~\ref{fig:model}, facilitating a more comprehensive understanding of the concepts and processes involved.
Secondly, Google Colab offers access to free GPU resources. This is a significant advantage for our participants as the computational power of GPUs can greatly expedite the execution of AI models, ensuring that experiments are completed within a reasonable time frame.
Furthermore, Google Colab's cloud-based nature eliminates the need for complex setups on personal machines, lowering the entry barrier for participants. This easy access, combined with the platform's robust functionality, makes Google Colab an ideal tool for our hands-on experiments in AI education.

\begin{figure}[h]
    \centering
    \includegraphics[width=0.48\textwidth]{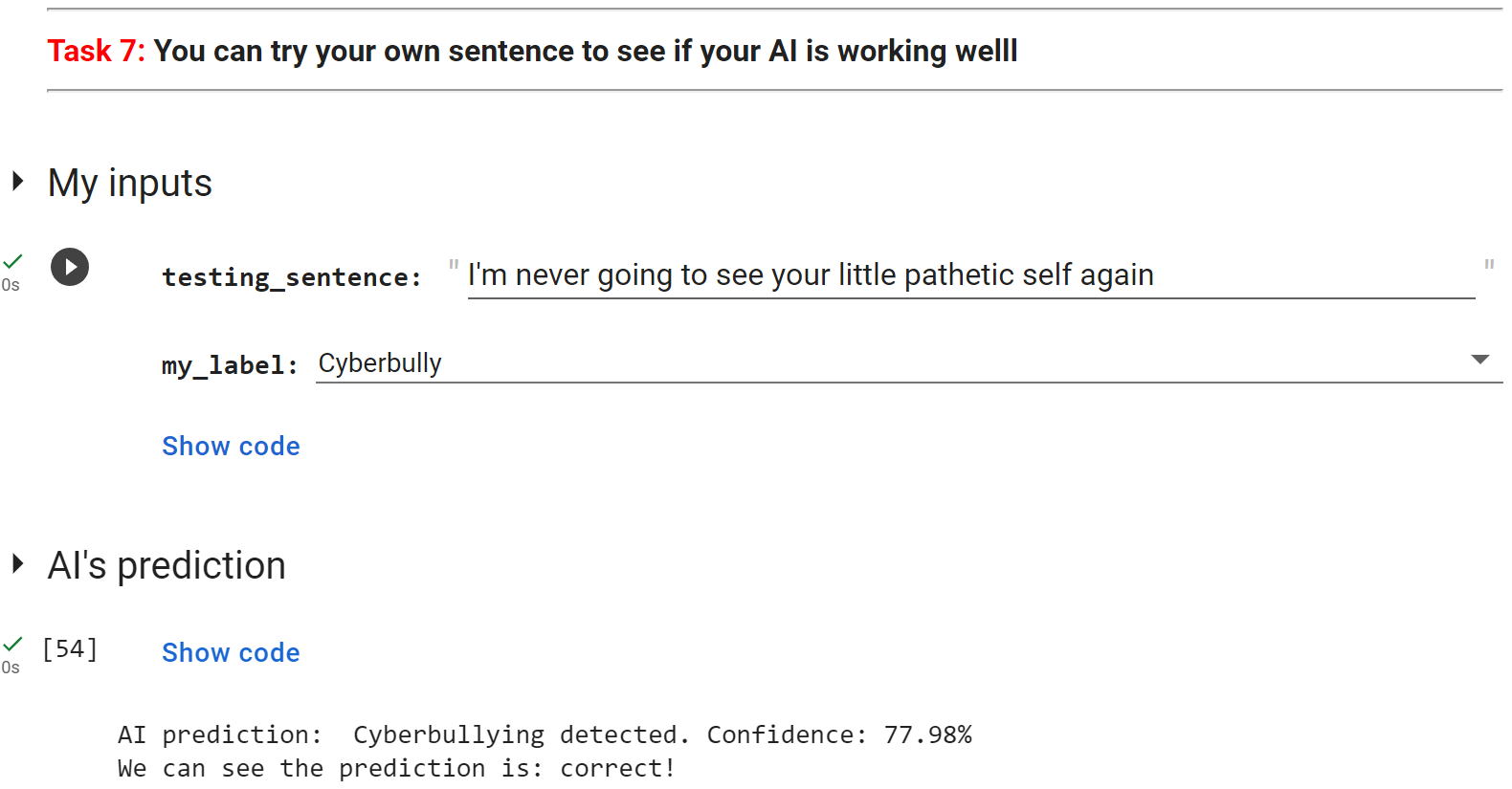}
    \caption{One example of a lab activity}
    \label{fig:task}
\end{figure}
In our labs, we have meticulously curated post-lab assignments that not only enhance the engagement factor but also deepen students' comprehension of the material. For example, as depicted in Figure~\ref{fig:task}, students are tasked with using arbitrary input statements to test their AI's capability to recognize cyberharassment content. 
In the Google Colab platform, the correct execution of a cell could depend on the execution of the previous cell, and students are made aware of this, which is also in the lab manual. In the example activity shown in Figure~\ref{fig:task}, students must complete all prior steps to access the developed AI model for cyberharassment detection. 
\begin{figure}[h]
    \centering
    \includegraphics[width=0.48\textwidth]{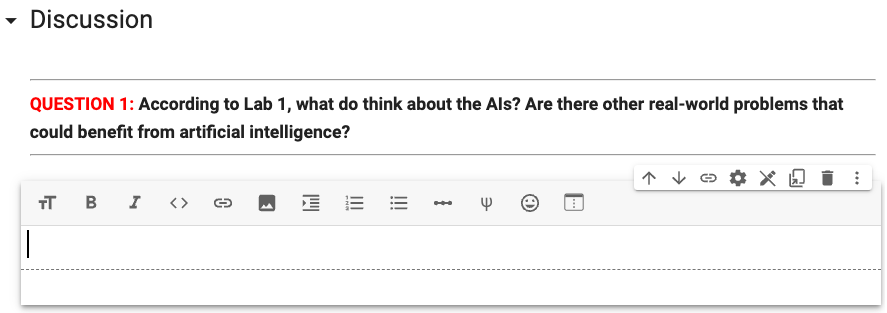}
    \caption{One example of Lab 1's discussion questions}
    \label{fig:question}
\end{figure}
Additionally, we have formulated post-lesson discussion questions. Figure~\ref {fig:question} depicts an example question. We aim to stimulate students' critical thinking by encouraging them to consider other real-world problems that could be alleviated through AI.

\subsection{Course Structure}
The Sociology program at North Carolina A\&T State University includes courses on Social Statistics, parts 1 and 2. Students from Social Statistics 1 were included in this intervention. Students in Social Statistics 1 learn how to interpret and describe data. Students are exposed to topics such as distributions; descriptive statistics (e.g., measures of central tendency and dispersion); statistical null-hypothesis testing, and independent and dependent samples t-tests. They also learn the basic operation of the statistical computing software program SPSS. Throughout the course, students learn the value statistical analysis offers to our attempts to address real-world social problems. 
This course runs 15 weeks and includes lecture and lab time for analysis. Students attend two 75-minute classes each week. Lectures involve definitions and the demonstration of practice problems. Lab time involves hands-on computing using SPSS. In Table \ref{table:demographics}, we report some basic demographics of the students in the Social Statistics 1 class in the Spring and Fall semesters. 

\begin{table}[!t]
\centering
\begin{tabular}{lcc}\toprule
& \multicolumn{1}{c}{Spring 2022} & \multicolumn{1}{c}{Fall 2022}
\\\cmidrule(lr){2-2}\cmidrule(lr){3-3}
Demographic  & Percent & Percent \\\midrule
Male  & 23.81\% & 14.29\% \\
Female & 71.43\% & 85.71\% \\
Freshman & 23.81\% & 4.76\% \\
Sophomore & 33.33\% & 66.67\% \\
Junior & 38.10\% & 9.52\% \\
Senior & 4.76\% & 14.29\% \\
Graduate & 0.00\% & 4.76\% \\\bottomrule
\end{tabular}
\caption{Demographics of students who participated in the Social Statistics 1 course in the Spring and Fall 2022 semesters.}
\label{table:demographics}
\end{table}

\section{Methods}
\subsection{Survey Design}
To evaluate the impact of our lab on students, we designed two surveys (pre-survey and post-survey) to collect quantitative and qualitative data from the students. Students rate their knowledge about AI and machine learning methods for cyberharassment detection. For both surveys, students rated their knowledge using a 5-point scale, with 1 representing "Proficient or strongly agree" and 5 representing "None or strongly disagree". In the post-survey, students also rate the lab from different perspectives. The post-survey also includes three open questions to understand what helped the students the most in understanding the concepts taught, the difficulties faced in using the lab, and any suggestions for improving the learning experience of the lab.

\subsection{Data Collection}
Before our data collection, we obtained our institution's Institution Review Board (IRB) approval. All the students were notified that the survey would remain confidential and only the research team could view the data. Students were also informed that their participation was optional. Before the class, students filled out the pre-survey to evaluate their knowledge level, and after the lab, students filled the post-survey. Our surveys are conducted using the Qualtrics platform~\cite{Qualitrics}.

\subsection{Data Analysis}
To analyze the collected data, we first compared the average knowledge score between pre-survey and post-survey. Then we used the sample t-test to determine the existence of statistical significance between the observed differences. We use a significance level of 0.05. If the p-value of the t-test is less than 0.05, we conclude a significant difference exists. Scipy, an open-source Python library for scientific computing, was used to analyze the collected data.

\begin{table}[!t]

	\centering
	\resizebox{8.5cm}{!}{
		\begin{tabular}
		{  >{\arraybackslash}m{4.8cm}|
		    >{\centering\arraybackslash}m{0.7cm}|
      >{\centering\arraybackslash}m{0.7cm}
		}

        \hline

1 = Proficient, 5 = None
&  Pre  
&  Post 
\\\hline
Automated cyberharassment detection.                                       &  3.95 
&  3.56 
\\\hline
State-of-The-Art 
cyberharassment detectors.
& 4.33
& 3.67
\\\hline
How machine learning works.
& 4.10  
& 3.89
\\\hline

{ The lab engaged me in learning the topic of AI Driven Socially-Relevant Cybersecurity.} & { - }     & { 2.78} 
\\\hline

{ I enjoyed the learning experience of this lab(s).}                                       & { - }     & { 2.89} 
\\\hline

{ I think the learning experience with the lab(s) is effective.}                           & { - }     & { 2.78} 
\\\hline

{ I am satisfied with the level of effort the lab requires for learning this topic.}      & { - }     & { 2.78} 
\\\hline

{ After using the lab(s), I have better understanding about the concepts learned.}         & { - }     & {3.44}               
\\\hline
		\end{tabular} 
	}
 \caption{Spring 2022 semester survey results.}
 \label{table:spring}
\end{table}

\begin{table}[!t]
	\centering
	\vspace{0pt}
	\resizebox{8.5cm}{!}{
		\begin{tabular}
		{  >{\arraybackslash}m{4.8cm}|
		    >{\centering\arraybackslash}m{0.8cm}|
      >{\centering\arraybackslash}m{0.8cm}
		}

        \hline

1 = Proficient, 5 = None
&  Pre  
&  Post 
\\\hline
Automated cyberharassment detection.                          &  4.38
& 3.31
\\\hline
State-of-The-Art 
cyberharassment detectors.
& 4.71
& 3.44
\\\hline
How machine learning works.
& 4.43
& 3.13
\\\hline

{ The lab engaged me in learning the topic of AI Driven Socially-Relevant Cybersecurity.} & { - }     & { 2.56} 
\\\hline

{ I enjoyed the learning experience of this lab(s).}                                       & { - }     & { 2.44} 
\\\hline

{ I think the learning experience with the lab(s) is effective.}                           & { - }     & { 2.13} 
\\\hline

{ I am satisfied with the level of effort the lab requires for learning this topic.}      & { - }     & { 2.56} 
\\\hline

{ After using the lab(s), I have better understanding about the concepts learned.}         & { - }     & {2.31}               
\\\hline
		\end{tabular} 
	}
 \caption{Fall 2022 semester survey results.}
 \label{table:fall}
\end{table}

\section{Results AND Discussion}

We analyze the data for the Spring 2022 and Fall 2022 semesters and compare the learning experiences of both semesters because the student feedback from the Fall 2022 semester, as discussed in the lesson learned section \ref{lesson_learned}, provided data and insights that were used to inform and refine the lab in the Spring 2022 semester. 
Before introducing the labs to the students, they were invited to complete a pre-questionnaire, and after completing the labs, they were invited to complete a post-questionnaire. The pre and post-questionnaire contained eleven questions in total. The first three questions were asked in both the pre and post-questionnaire. The remaining eight questions were only asked in the post-questionnaire in addition to the first three questions. Of the eight questions, the last three were open-ended questions. In the initial data analysis, we focus on the first three questions in the pre-and-post questionnaires administered to the students. In the final data analysis, we focus on only the last eight questions in the post-questionnaire that the students completed after completing the labs to gauge their understanding and perception of the lab. 
The data analysis focused on survey participation, knowledge in automated cyberharassment detection, evaluation of ML classifiers, current state-of-the-art cyberharassment systems, how ML works, general learning experience and engagement of the labs, and student qualitative feedback. Tables~\ref{table:spring} and \ref{table:fall} show the Spring 2022 and Fall 2022 survey results.  

\subsection{Survey Participation}

The survey was presented to students in the Spring 2022 and Fall 2022 semesters. Of the 21 students enrolled in the course in Spring 2022, 21 (100\%) students completed the pre-survey, and 9 (43.9\%) students completed the post-survey. In the Fall 2022 semester, 21 students were enrolled in the class, of which all students completed the pre-survey, and 16 (80\%) students completed the surveys—showing that the Fall 2022 session of the course had more student participation than when the lab was first introduced in Spring 2022. 

\subsection{Cyberharassment Detection}

During the pre-survey of Spring 2022, 19.05\% of the students rated their knowledge or skills in automated cyberharassment detection in the \textit{Proficient} and \textit{Good} category, about 14.29\% of the students rated themselves as having moderate knowledge or skill, and more than half (66.67\%) of the students rated their knowledge or skills in the \textit{A little} and \textit{None} category. After the post-survey, 33.33\% of the students rated their knowledge or skills in the \textit{Proficient} and \textit{Good} category, 0\% rated themselves as moderate, and 66.66\% rated their knowledge in the \textit{A little} and \textit{None} category.

In the pre-survey Fall 2022, 4.76\% of the students rated their knowledge in the \textit{Proficient} and \textit{Good} category, 19.05\% as having moderate skill or knowledge, and 76.19\% rated their knowledge or skills in the \textit{A little} and \textit{None} category. On the other hand, in the post-survey of Fall 2022, 25\% of the students rated their knowledge in the \textit{Proficient} and \textit{Good} categories, 25\% rated themselves as moderate, and 50\% rated their knowledge in the \textit{A little} and \textit{None} category. 

Most students had little knowledge of automated cyberharassment detection before enrolling in the class and participating in the lab. After completing the lab and comparing the means of the pre-survey and post-survey of Spring 2022 as shown in Table~\ref{table:spring}, there is no significant difference ($p > 0.05$) in the knowledge of automated cyberharassment detection. For the Fall 2022 semester, there is a significant difference ($p < 0.05$) in the knowledge of automated cyberharassment detection as indicated in Table~\ref{table:fall}. Indicating that the improvements made after the Spring 2020 semester helped improve students' knowledge or skill in the Fall 2022 semester.

\subsection{State of the Art Cyberharassment Systems}
We determined the knowledge or skills of students in the state-of-the-art cyberharassment systems before and after the lab. During the pre-survey of Spring 2022, 4.76\% of the students rated their knowledge or skills in state-of-the-art systems in the \textit{Proficient} and \textit{Good} categories, about 19.05\% of the students rated themselves as having moderate knowledge or skill, and more than half (76.19\%) of the students rated their knowledge or skills in the \textit{A little} and \textit{None} category. After the post-survey, 33.33\% of the students rated their knowledge or skills in the \textit{Proficient} and \textit{Good} category, 0\% rated themselves as moderate, and 66.66\% rated their knowledge in the \textit{A little} and \textit{None} category.

In the pre-survey Fall 2022, 4.76\% of the students rated their knowledge in the \textit{Proficient} and \textit{Good} categories, 4.76\% as having moderate skill or knowledge, and 90.47\% rated their knowledge or skills in the \textit{A little} and \textit{None} category. On the other hand, in the post-survey of Fall 2022, 18.75\% of the students rated their knowledge in the \textit{Proficient} and \textit{Good} categories, 25\% rated themselves as moderate, and 56.25\% rated their knowledge in the \textit{A little} and \textit{None} category. 

Many students had little knowledge of the state-of-the-art cyberharassment system before the lab. After completing the lab and comparing the means of the pre-survey and post-survey of Spring 2022, there is no significant difference ($p > 0.05$) in the knowledge of state-of-the-art systems. For the Fall 2022 semester, there is a significant difference ($p < 0.05$) in the knowledge of state-of-the-art systems. Similar to the knowledge of cyberharassment detection, the improvements made after the Spring 2020 semester helped in improving students' knowledge or skill in the Fall 2022 semester.

\subsection{How Machine Learning Works}
We also determined if the students knew how machine learning worked before the lab and if they learned how machine learning worked after the lab. From the pre-survey of Spring 2022, results show that 9.52\% of the students rated their knowledge in how ML worked in the \textit{Proficient} and \textit{Good} categories, about 28.57\% of the students rated themselves as having moderate knowledge, and more than half (61.9\%) of the students rated their knowledge or skills in the \textit{A little} and \textit{None} category. After the post-survey, 22.22\% of the students rated their knowledge in the \textit{Proficient} and \textit{Good} category, 0\% rated themselves as moderate, and 77.77\% rated their knowledge in the \textit{A little} and \textit{None} category.

In the pre-survey Fall 2022, 4.76\% of the students rated their knowledge in the \textit{Proficient} and \textit{Good} categories, 14.29\% as having moderate skill or knowledge, and 80.96\% rated their knowledge or skills in the \textit{A little} and \textit{None} category. On the other hand, in the post-survey of Fall 2022, 26.66\% of the students rated their knowledge in the \textit{Proficient} and \textit{Good} categories, 33.33\% rated themselves as moderate, and 40\% rated their knowledge in the \textit{A little} and \textit{None} category. 

In both semesters, more students had little knowledge of how ML worked before the lab. After completing the lab and comparing the means of the pre-survey and post-survey of Spring 2022, there is no significant difference ($p > 0.05$) in the knowledge of how ML worked. For the Fall 2022 semester, there is a significant difference ($p < 0.05$) in the knowledge of how ML worked. The lab refinements made after the Spring 2020 semester helped improve students' knowledge or skill in the Fall 2022 semester.

\subsection{Lab Engagement}
The students responded positively to how the lab engaged them in learning about AI-driven socially relevant cybersecurity. In the Spring of 2022, 44.44\% of the students rated how engaging the lab was in the \textit{Strongly agree} and \textit{Somewhat agree} categories, 22.22\% rated engagement as moderate, and 33.33\% rated engagement in the \textit{Somewhat disagree} and \textit{Strongly disagree} category. In the Fall of 2022, more than half (62.5\%) of the students rated how engaging the lab was in the \textit{Strongly agree} and \textit{Somewhat agree} categories, 12.50\% rated engagement as moderate, and 25\% rated engagement in the \textit{Somewhat disagree} and \textit{Strongly disagree} category. 

The mean response by students in the Spring and Fall semesters was 2.78 and 2.56, respectively. The mean values are between the \textit{Somewhat agree} and \textit{Neither agree nor disagree} with the mean value of Fall closer to \textit{Somewhat agree} than the Spring mean, indicating that the students' lab engagement was positive and not very negative. 

\subsection{Learning Experience}
The student's response to the overall learning experience and the effectiveness of the experience was somewhat positive. In the Spring of 2022, 33.33\% of the students rated the learning experience in the \textit{Strongly agree} and \textit{Somewhat agree} categories, 44.44\% rated the learning experience as moderate, and 22.22\% rated the learning experience in the \textit{Somewhat disagree} and \textit{Strongly disagree} category. For the effectiveness of the learning experience, 44.44\% of the students rated the effectiveness in the \textit{Strongly agree} and \textit{Somewhat agree} category, 33.33\% rated effectiveness as moderate, and 22.22\% rated learning experience in the \textit{Somewhat disagree} and \textit{Strongly disagree} category.

In the Fall of 2022, more than half (68.75)\% of the students rated the learning experience in the \textit{Strongly agree} and \textit{Somewhat agree} categories, 12.50\% rated learning experience as moderate, and 18.75\% rated learning experience in the \textit{Somewhat disagree} and \textit{Strongly disagree} category. For the effectiveness of the learning experience, 43.75\% of the students rated the effectiveness in the \textit{Strongly agree} and \textit{Somewhat agree} category, 37.50\% rated effectiveness as moderate, and 18.75\% rated learning experience in the \textit{Somewhat disagree} and \textit{Strongly disagree} category. 

For the overall learning experience, the mean response by students in the Spring and Fall semesters was 2.89 and 2.44, respectively. The mean value of the Fall semester is closer to \textit{Somewhat agree}, indicating that the students had a more positive learning experience in the Fall semester. For the effectiveness of the learning experience, the mean response from students in the Spring and Fall semesters was 2.78 and 2.13, respectively. Similar to the learning experience, the mean value of the Fall semester is closer to \textit{Somewhat agree}, indicating that in the Fall semester, the effectiveness of the learning experience was more positive for the students. 

\subsection{Lab Difficulty}
We gauged the difficulty of the lab for the students, and the response indicated that the students neither agreed nor disagreed that the labs were demanding. In the Spring of 2022, 44.44\% of the students rated the level of effort required to learn the topic in the \textit{Strongly agree} and \textit{Somewhat agree} categories, 22.22\% rated engagement as moderate, and 33.33\% rated learning effort in the \textit{Somewhat disagree} and \textit{Strongly disagree} category. In the Fall of 2022, more than half (62.5\%) of the students rated how engaging the lab was in the \textit{Strongly agree} and \textit{Somewhat agree} categories, 25\% rated learning effort as moderate, and 12.50\% rated learning effort in the \textit{Somewhat disagree} and \textit{Strongly disagree} category. 

The mean response by students in the Spring and Fall semesters was 2.78 and 2.56, respectively. The mean values are between the \textit{Somewhat agree} and \textit{Neither agree nor disagree} with the mean value of Fall closer to \textit{Somewhat agree} than the Spring mean, indicating that the level of effort required to learn the topic by the students was better in the Fall than in Spring even though the students neither agree nor disagree.  

\subsection{Understanding of Concepts}
To understand whether the students better understood the concepts learned, we determined students' grasp of the concepts introduced in the lab. In the Spring of 2022, 22.22\% of the students rated understanding of concepts in the \textit{Strongly agree} and \textit{Somewhat agree} categories, 33.33\% rated understanding of concepts as moderate, and 44.44\% rated understanding of concepts in the \textit{Somewhat disagree} and \textit{Strongly disagree} category. In the Fall of 2022, 43.75\% of the students rated understanding of concepts in the \textit{Strongly agree} and \textit{Somewhat agree} categories, 25\% rated understanding of concepts as moderate, and 31.25\% rated learning effort in the \textit{Somewhat disagree} and \textit{Strongly disagree} category. 

The mean response by students in the Spring and Fall semesters was 3.44 and 2.31, respectively. The mean value for the Spring semester is between \textit{Neither agree nor disagree} and \textit{Somewhat disagree}. For the Fall semester, the mean value is between \textit{Somewhat agree} and \textit{Neither agree nor disagree}. These values indicate that in the Spring semester, students understanding of concepts learned was closer to negative (Somewhat disagree). However, in the Fall semester, students better understood the concepts learned. 

\subsection{Qualitative Feedback}
Our last three survey questions were open-ended questions about what has been the most helpful for learning, what has caused the most difficulty using the lab, and how the lab can be improved. We used these as the qualitative data source, providing insights into students' perceptions and preferences. In general, the students understood the purpose of the lab and cyberharassment, found the terminology confusing, and wanted the lab to be more fun and engaging. In the Spring 2022 semester, in response to \say{What has been most helpful for your learning in using the lab so far}. Notable student responses were: \textit{\say{I understand the purpose of cyberbullying and its purpose and how it is designed to be successful}} and \textit{\say{The guest speakers coming in to help}}. For the Fall 2022 semester, the notable student responses were: \textit{\say{The most helpful part for my learning has been the hands-on activity, being able to ask questions while going through the work and having a guest speaker gave some new insight.}}, \textit{\say{I learned a lot about cyber bullying that I did not know about and the different forms it can come in.}}, \textit{\say{The lab was helpful with detecting cyber harassment.}}, \textit{\say{It was nice knowing the set up and watching the steps be performed}}, and \textit{\say{The videos and zoom call}}. 

For the question \say{In terms of your learning, what has caused you the most difficulty in using the lab so far}. The notable student responses in the Spring 2022 semester were: \textit{\say{I had a hard time understanding how to actually complete on my own}}, \textit{\say{The terminology}}, \textit{\say{I could not stay focus and lacked engagement}}, and \textit{\say{Being online.}} Notable responses in the Fall semester were: \textit{\say{It was difficult that when there was a troubleshooting error that I could not walk through it with someone like I could in person.}}, \textit{\say{The most difficult part is not usage of the lab itself; it is remembering certain aspects of what to do and what to look for when in the lab.}}, \textit{\say{The most difficulty experienced in the labs is facing errors and technical difficulties.}}, \textit{\say{I think the directions were hard to follow because the instructor seemed like he assumed we knew something about this content.}}, and \textit{\say{Fully understanding what was being done in the lab was the most difficult.}}

Finally, for the question \say{What suggestion(s) can you make that would enhance your learning experience with the lab}. Notable student responses in the Spring 2022 semester were: \textit{\say{Make the lab more engaging/fun}}, \textit{\say{Break down steps on how to actually complete the activity}}, \textit{\say{I would say, trying using a different online platform for this lab, to make everything a little bit easier for students to understand.}}, \textit{\say{Being in person}}, and \textit{\say{Better terminology}}. In the Fall 2022 semester, notable responses were: \textit{\say{The instructor was helpful it’s just hard to learn over the computer such a difficult thing to do.}}, \textit{\say{I cannot think of anything at the moment. I really enjoyed learning about this lab and how it worked.}}, \textit{\say{An in person option}}, \textit{\say{provide a tutorial video.}}, \textit{\say{I would say slowing down the directions.}}, and \textit{\say{Teach more about how to face technical difficulties.}}

\subsection{Lesson Learned}\label{lesson_learned}
The authors learned the following lessons in implementing a cloud-based laboratory experience. We outline tips for developing a cloud-based laboratory for teaching AI socially relevant cybersecurity. 
\subsubsection{Code Dilution}
The lab is implemented on Google Colab, Google's cloud-based jupyter notebook platform. The initial implementation of the lab on Google Colab presented the students with the raw code implementation of cyberharassment detection using PyTorch, an open-source Python framework for developing ML, especially deep learning systems. From the Spring 2022 feedback from the students, we observed that the students were not positive towards the code implementation since they have very little programming experience. This frustrated students and slowed interest and learning. With this knowledge, in preparation for the Fall 2022 semester, we improved the learning experience by re-implementing the lab with the code hidden to enable the students to think about social issues and focus on understanding how AI can be used to approach social issues such as cyberharassment.  

\subsubsection{Lab Instructions}
Developing the lab instructions in the lab manual is crucial to enhancing the student experience. If the lab instructions are not very detailed, with step-by-step instructions on how to complete the lab activities, the students struggle with understanding and completing the lab independently. From the Spring 2022 student feedback, students complained that the instruction manual needed more detail and felt the instructors assumed they were familiar with AI and AI terminology. In the Fall 2022 semester, we improved the lab manual by toning down the terminology and providing more step-by-step descriptions so the students could complete the labs independently and understand the purpose and rationale behind each step. 

\subsubsection{Pre-lab Lecture}
Before allowing the students to complete the lab independently, we prepared lecture slides about the lab and introduced them to the problem and the activities they would be completing. The students found this pre-lab lecture particularly helpful. 

Other best practices include recording the pre-lab lecture so that the students can refer back to it when working on the lab activities, anticipating possible technical difficulties, and including steps to solve the issues in the lab manual. Having an in-person option where the students can complete the labs during class could help with engagement and technical issues they might encounter. 

\section{Limitations}
Our work has limitations. First, our study is focused on one public institution in the United States, limiting our findings' generalizability. Second, the generalizability of our findings is also limited by the focus of our study on one non-computing program - Social Statistics, excluding other non-computing programs. Third, the number of participants in our study further limits our conclusions. Finally, the current lab only focuses on the general concepts of AI and how it can be utilized to address social issues such as cyberharassment. It does not cover other areas of AI, such as how adversarial attacks can fool cyberharassment models, multi-modal cyberharassment detection, and the issue of fairness and bias in cyberharassment models. 
\section{Conclusion and Future Work}
We have developed an AI socially relevant cybersecurity lab for cyberharassment detection for non-computing students. We introduced a cyberharassment detection lab's development, implementation, and assessment. The development process has been guided by the learning objective of introducing a hands-on experiential lab that will increase general awareness of socially relevant cybersecurity and AI and is suitable for teaching AI socially relevant cybersecurity to non-computing students. Our lab offers meaningful experiential learning opportunities that allow students to work on real-world social issues such as cyberharassment. After incorporating student feedback in the redesign of the lab used in the Fall semester, the knowledge or skills of most students in automated cyberharassment detection and how ML works improved significantly compared to the Spring semester. Also, students found the detection of cyberharassment helpful and understood the purpose of using AI for social issues. We plan on continuing to refine the lab and use the knowledge gained in developing four labs currently under development that cover multi-modal (text and image) cyberharassment detection, adversarial attacks on cyberharassment systems, bias mitigation in cyberharassment systems, and the use of generative AI models such as ChatGPT for cyberharassment detection. Additionally, we plan on developing these labs for computer science and engineering students in the future.

\section*{Acknowledgment}
The work was supported by National Science Foundation (NSF) under the Grant No. 2239605, 2228616 and 2114920.

\bibliography{ref}

\end{document}